\documentclass[usenatbib]{mn2e}
\usepackage{graphicx}

\newcommand\araa{{ARA\&A}}
\newcommand\apj{{ApJ}}
\newcommand\aap{{A\&A}}
\newcommand\mnras{{MNRAS}}

\def\omega0{\Omega_{\\rm m,0}}

\def\kms{\,\rm km\,{s}^{-1}}
\def\kpc{\,\rm kpc}
\def\mpc{\,\rm Mpc}
\def\LCDM{\Lambda{\rm CDM}}
\def\Ds{D_{\rm s}}

\def\zs{z_{\rm s}}
\def\zl{z_{\rm l}}
\def\beq{\begin{equation}}
\def\eeq{\end{equation}}

\date{
Accepted 2006 August 7.
Received 2006 August 4;
in original form 2006 July 5}

\pubyear{2006}

\title{The giant arc statistics in the three year WMAP cosmological model}
\author[Li et al.]{
G. L. Li$^{1,2}$\thanks{E-mail: {\tt lgl@shao.ac.cn}},
S. Mao$^3$,
Y.P. Jing$^{1,2}$,
H.J. Mo$^{4}$,
L. Gao$^{5}$,
W.P. Lin$^{1,2}$ 
\\
$^{1}$ Shanghai Astronomical Observatory, Nandan Road 80, Shanghai 200030, China \\
$^2$ Joint Institute for Galaxy and Cosmology (JOINGC) of SHAO and USTC \\
$^3$ University of Manchester, Jodrell Bank Observatory, Macclesfield, Cheshire SK11 9DL, U.K. \\
$^4$ Department of Astronomy, University of Massachusetts, Amherst MA 01003-9305, USA \\
$^5$ Institute for Computational Cosmology, Physics Department, Durham, DH1 3LE, U.K.
}

\begin{document}
\maketitle

\begin{abstract}
We use high-resolution $N$-body simulations to investigate the
optical depth of giant arcs with length-to-width ratio larger than 7.5
and 10 in the `standard' $\LCDM$ model with $\sigma_8=0.9$ and
$\Omega_{\rm m,0}=0.3$ and a model based on three-year Wilkinson
Microwave Anisotropy Probe (WMAP) data. We find that, in dark-matter
only simulations, the lensing probability in the three-year WMAP
model (with $\sigma_8=0.74$ and $\Omega_{\rm m,0}=0.238)$
decreases by a factor of $\sim 6$ compared with that in the `standard'
$\LCDM$ model. The effects of baryonic cooling, star formation and
feedbacks are uncertain, but we argue that baryons will only increase
the  the lensing cross-section by a moderate factor, $\sim 2$. We
conclude that the low central value of $\sigma_8$ and $\Omega_{\rm m,0}$
 preferred by the WMAP three-year data may be too low to be compatible with 
observations if conventional assumptions of the background source
population are correct.
\end{abstract}

\section{INTRODUCTION}

In the hierarchical scenario of structure formation,
structures in the universe are assumed to have grown 
from tiny quantum fluctuations generated during an inflation 
period through gravitational instability. Inflationary
models predict that the initial fluctuations are Gaussian and
have a roughly scale-free power-spectrum: $P(k) \propto k^n$
with $n \sim 1$. The amplitude of the power spectrum, for which 
a reliable prediction from inflationary theory is still lacking, 
has to be determined from observations. This amplitude is 
usually represented by a quantity, $\sigma_8$, which is  
the {\it RMS} of the linear density perturbations within 
a spherical window of radius $8h^{-1}\mpc$, where $h$ the Hubble 
constant in units of 100$\kms\mpc^{-1}$. 

The linear power spectrum can be probed with a variety of
observations, ranging from  abundances of clusters, weak
gravitational lensing, large-scale structure in galaxy distribution,
galaxy motions, Lyman-$\alpha$ forests, and most importantly,  the
temperature fluctuations in the cosmic microwave background (see
Spergel et al. 2006, and references therein). Recent results from the
three-year data of the Wilkinson Microwave Anisotropy Probe (WMAP)
alone prefers a $\sigma_8$ value of  $0.74^{+0.06}_{-0.05}$. Although
the errorbar is still quite large, the median value is lower than the
value, $\sigma_8=0.9$, adopted in the `standard' $\LCDM$ model. The
lower value of $\sigma_8$ is favored by a number of other observations
related to the clustering of galaxies (e.g. Jing, Mo \& B\"orner 1998; 
Yang et al. 2004, 2005; van den Bosch, Mo \& Yang 2003; 
Tinker et al. 2006; S\'{a}nchez et al. 2006), but weak lensing results
(e.g. \citealt{Bacon02,Van Waerbeke02} ) and data on Lyman alpha
forests (Viel, Haehnelt \& Lewis 2006; Seljak, Slosar \& McDonald
2006) seem to prefer a higher value. 

As the largest bound structures in the universe, clusters of galaxies
provide a sensitive probe for $\sigma_8$, because their abundance in
the universe depends sensitively on the amplitude of the density
perturbations (and also the matter content in the universe,
$\Omega_{\rm m,0}$). The abundance of clusters can be determined from
X-ray (e.g. \citealt{Rosati02,Pop04, RB02, R06}) and optical surveys
(e.g. \citealt{Gladders06}). However, in order to use the observed
abundance to constrain model of structure formation, one has to know
the masses of these objects accurately. This is in general difficult
to do.  For example, X-ray studies usually assume hydrostatic
  equilibrium in order to derive the cluster mass, which may be
  invalid for many (such as merging) clusters (e.g. Gao \& White
2006). Another probe of the abundance of clusters is  giant arcs
that are produced when background galaxies are tangentially stretched
by foreground clusters (e.g. \citealt{Luppino99, ZG03, Gladders03, Sand05}). 
A number of recent investigations made comparisons between the 
observed abundance of giant arcs with that expected in the 
`standard' $\LCDM$ model with $\sigma_8=0.9$ (\citealt{Dalal04}; 
\citealt{Li05}; \citealt{Hen05}; \citealt{Horesh05}), 
finding agreement between model and observation 
(\citealt{Dalal04, Wambsganss04}). In light of the change in 
cosmological parameters preferred by  the WMAP three-year data, 
it becomes important to re-evaluate how the predicted lensing probability changes.

In this paper, we study the number of giant arcs expected in the new
cosmological model, using high resolution $N$-body simulations. We
compare the results with those obtained from similar simulations of
the standard $\Lambda$CDM model, and with current observations. The
plan of the paper is as follows. In \S2, we describe the simulations
we use and the analysis to predict giant arcs in the $N$-body
simulations. Our main results are presented in \S3, and we finish
with a discussion in \S4.

\section{Cosmological models, numerical simulations, and lensing method}

In this paper, we use two sets of simulations, one for the `standard'
$\LCDM$ model with $\sigma_8=0.9$, and the other for the cosmological
model given by the recent three-year WMAP data with
$\sigma_8=0.74$. For brevity, these two models will be referred to as
$\LCDM$0, and WMAP3, respectively, and the corresponding cosmological
parameters are:
\begin{enumerate}
\item $\LCDM$0: $\Omega_{\rm m,0}=0.3,\Omega_{\Lambda,0}=0.7,h=0.7,
     \sigma_8=0.9, n=1$;
\item WMAP3: $\Omega_{\rm m,0}=0.238,\Omega_{\Lambda,0}=0.762,
     h=0.73,\sigma_8=0.74, n=0.95$.
\end{enumerate}
The assumed initial transfer function in each model was generated
  with {\small CMBFAST} (Seljak \& Zaldarriaga 1996). Notice that both
  $\sigma_8$ and $\Omega_{\rm m,0}$ differ in these two models.

\begin{table}
\caption{Cosmological simulation parameters. The columns are 
cosmology, box size, number and mass of dark matter particles, and the softening length.}
\begin{tabular}{lcccc} 
\hline
model & box size & $N_{\rm DM}$ & $m_{\rm DM}$  &  softening \\
      & $(h^{-1}\mpc)$  &       &   $(h^{-1} M_\odot)$          &  ($h^{-1}$\kpc) \\\hline
$\LCDM$0 & 300 &  $512^3$  &  $1.67\times 10^{10}$   &  30  \\
WMAP3 & 300 &  $512^3$     & $1.32\times 10^{10}$  &  30  \\
\hline
\end{tabular}
\label{table:simu}
\end{table}

We use a vectorized-parallel ${\rm P^3M}$ code (Jing \& Suto 2002)
and a {\small PM-TREE} code -- GADGET2 \citep{Springel01, Springel05} to
simulate the structure formation in these models; the details of
the simulations are given in Table 1. Both are $N$-body simulations
which evolve $N_{\rm DM}=512^3$ dark matter particles in a large cubic
box with sidelength equal to $300 h^{-1} \mpc$. The $\LCDM$0
simulation was performed using the ${\rm P^3M}$ code  of Jing \& Suto
(2002) and has been used in \citet{Li05} to study the properties of
giant arcs. We refer the readers to that paper for the detail. 
The WMAP3 simulation is a new simulation carried out with GADGET2. 
With their large volumes, these two simulations sample the
corresponding cluster mass functions reasonably well, and they will 
be used to compare the optical depths of giant arcs in the $\LCDM$0 
and WMAP3 models.

For each simulation, we identify virialised dark matter halos using the 
friends-of-friends method with a linking length equal to 0.2 times the mean
particle separation. The halo mass $M$ is defined as the virial mass
within the virial radius according to the spherical collapse
model (Kitayama \& Suto 1996; Bryan \& Norman 1998; Jing \& Suto 2002).
Giant arcs are produced mainly by clusters of galaxies, and so we focus
on massive haloes with $M\ga 10^{14}h^{-1} M_\odot$.

For a given cluster, we calculate the smoothed surface density maps
using the method of \citet{Li06}. Specifically, for any line
of sight, we obtain the surface density on a $1024 \times 1024$ grid 
covering a square of (comoving) sidelength
of $4h^{-1}\mpc$ centered on each cluster. The projection depth is
chosen to be $4h^{-1}\mpc$. 
Particles
outside this cube and large-scale structures do not contribute 
significantly to the lensing
cross-section  (e.g.  \citealt{Li05}; \citealt{Hen05}). Our
projection and smoothing method uses a smoothed particle hydrodynamics (SPH) kernel to distribute
the particle mass on a 3D grid and then integrate along the line
of sight to obtain the surface density (see \citealt{Li06} for detail).
In this work, the number of neighbors used in the SPH smoothing kernel 
is fixed to be 32. Once a surface density map is obtained,  
we compute the cross-section of giant arc formation following 
the method given in \citet{Li05}. 
We consider seven source redshifts, $\zs=0.6, 1.0, 1.5, 2.0, 3.0, 4.0$ and 7.0. 
The background sources are assumed to be elliptical 
with random position angles and a fixed angular surface
area, $S_{\rm source}=\pi \times 0.5\arcsec^2$. The axis ratio is randomly drawn between 
0.5 and 1. This choice of axis ratio and source size is in good agreement with
the study of high-redshift galaxies using HST by \citet{Fer04}. 
We generate a large number of background sources within a rectangle box
with area $S_{\rm box}$. This rectangle is chosen to enclose all the high-magnification
regions that can potentially form giant arcs. The number of sources
generated is given by $n_{\rm source}=9S_{\rm box}/S_{\rm source}$.
For each source, ray-tracing is used to find the resulting image(s).
Giant arcs are identified as elongated 
images with length-to-width ($L/W$) ratio exceeding 7.5 or 10, the usual
criterion used to select giant arcs in observations.
We calculate the total cross sections of the top 200 most massive clusters in each 
simulation output and obtain the average cross section per unit comoving volume by:
\beq
\overline\sigma(\zl, \zs) = {\sum\sigma_i(\zl,\zs) \over V},
\eeq
where $\sigma_i(\zl, \zs)$ is the average cross-section of the 
three projections of the $i$-th 
cluster at redshift $\zl$, $\zs$ is the source redshift, and $V$ is
the comoving volume of the simulation box.
The optical depth can then be calculated as:
\beq \label{eq:tau}
\tau(\zs) = {{1\over{4\pi{\Ds}^2}} {\int_0}^{\zs}\, dz \, \overline\sigma(z,\zs)
(1+z)^3\, {dV_{\rm p}(z)\over dz}},
\eeq
where $\Ds$ is the angular diameter distance to the source, and
 $dV_{\rm p}(z)$ is 
the proper volume of a spherical shell with redshift from $z$ to $z+dz$.
The integration step size is the same as the redshift interval of
simulation output ($d z\approx 0.1$).

\section{Results}

\begin{figure}
{
 \centering
 \leavevmode 
\columnwidth=.48\columnwidth
 \includegraphics[width={\columnwidth}]{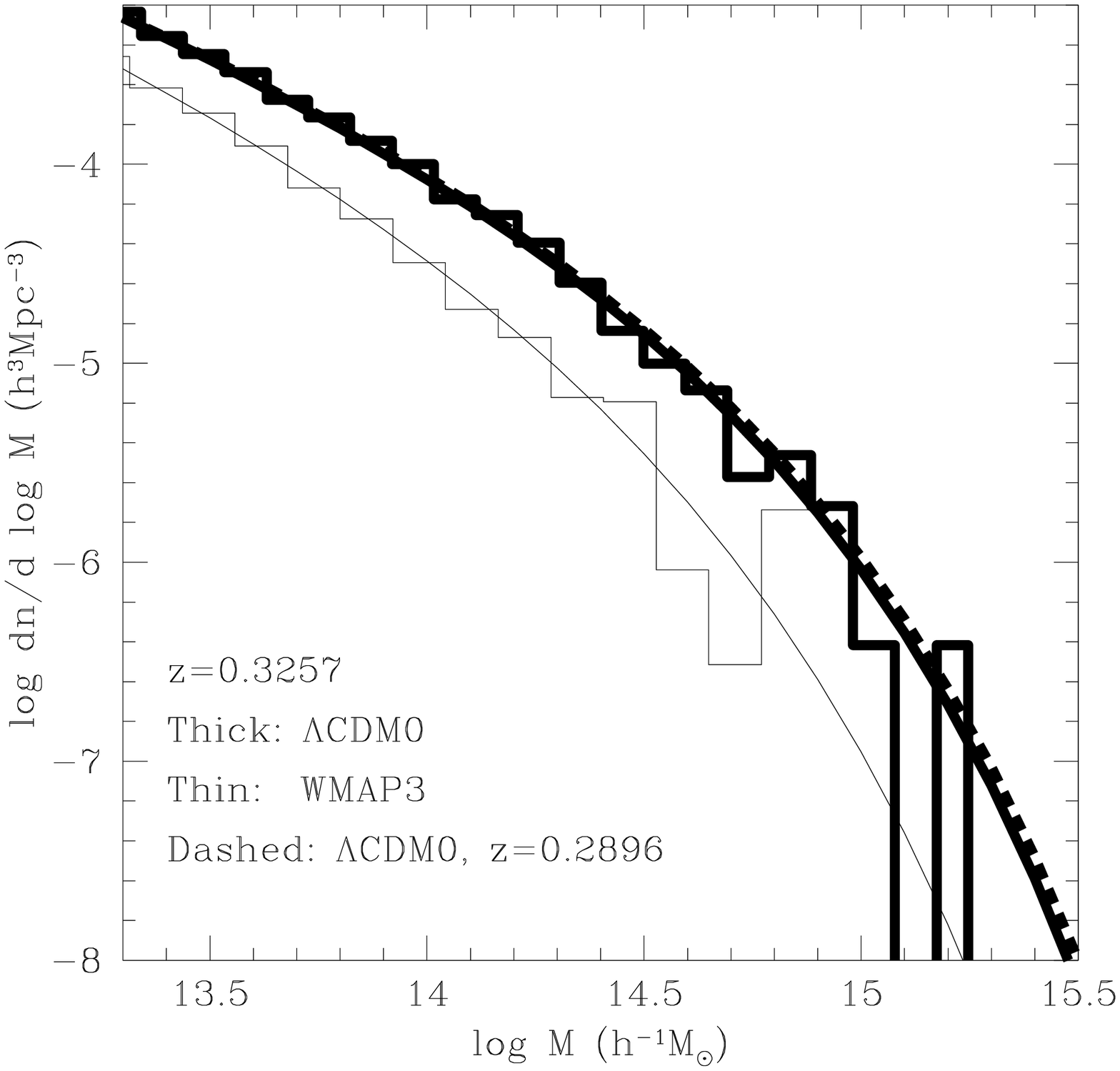}%
 \hfil
 \includegraphics[width={\columnwidth}]{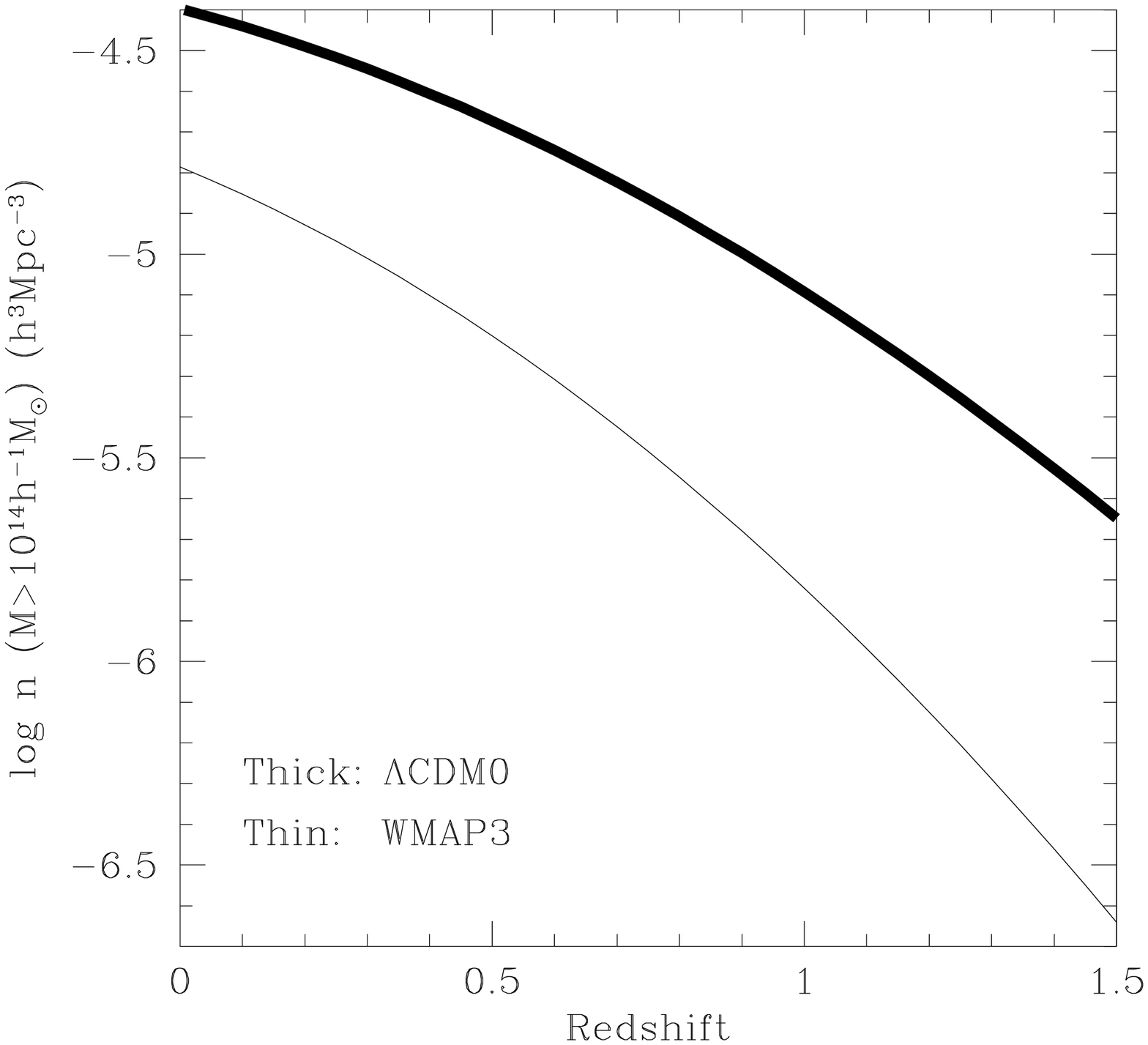}
}
\caption{The left panel shows the number density of halos in the
$\LCDM$0 (thick) and WMAP3 (thin) cosmological models (see text) at
redshift 0.33 obtained using the Sheth-Tormen mass function (2002). The thin solid histogram
shows the number density of haloes found in a simulation with box size
$300 h^{-1}\mpc$ in the WMAP3 model at $z=0.33$.
The thick solid histogram shows the number density of halos
in the $\LCDM$0 simulation at $z =0.2896$ (the corresponding
theoretical prediction is shown as the thick dashed line).
The right panel shows how the number density of haloes with mass $M>10^{14}h^{-1}
M_\odot$ evolves as a function of redshift in the two cosmologies.
}
\label{fig:massfunc}
\end{figure}

\begin{figure}
{
 \centering
 \leavevmode 
\columnwidth=.48\columnwidth
 \includegraphics[width={\columnwidth}]{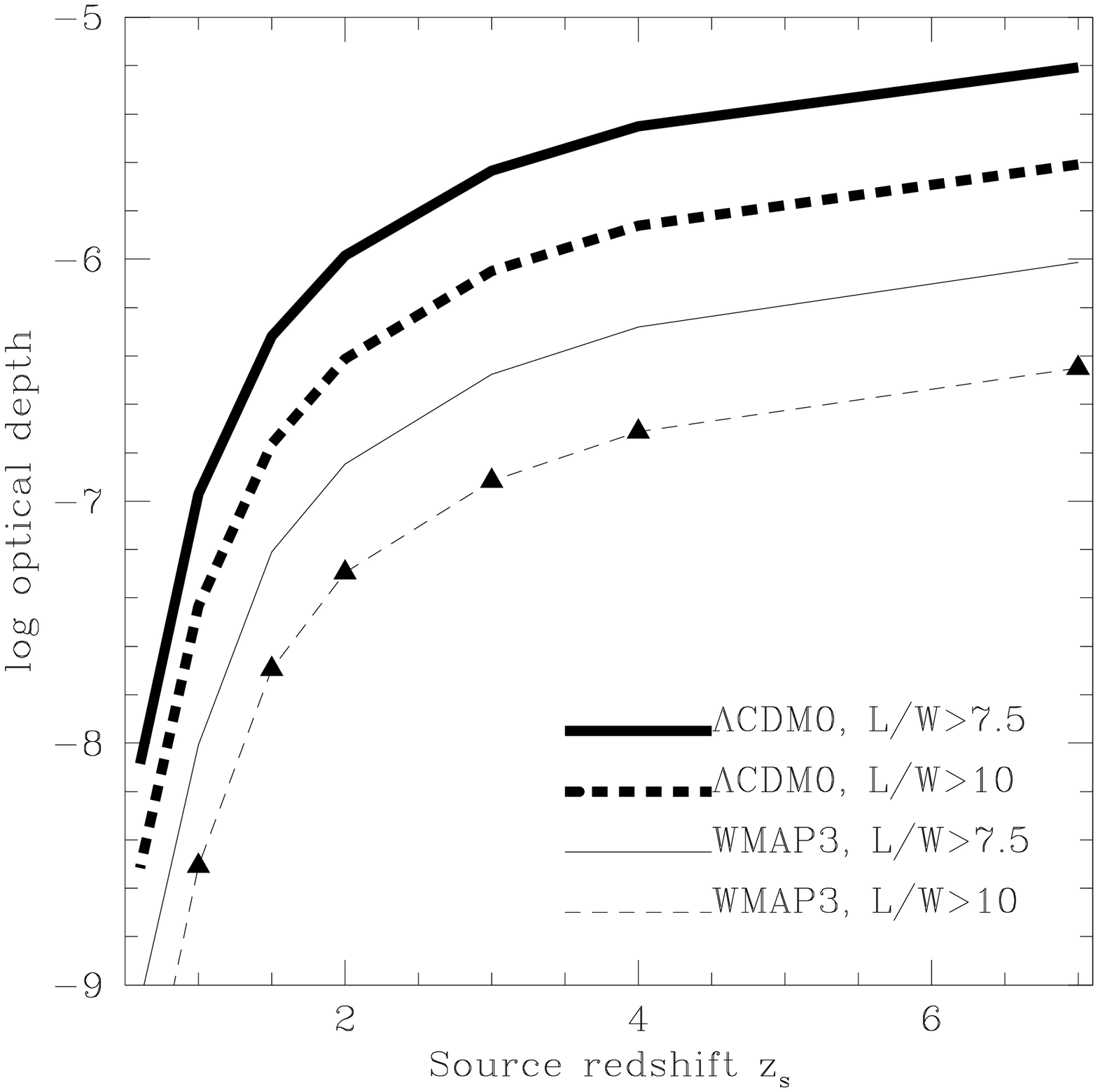}%
 \hfil
 \includegraphics[width={\columnwidth}]{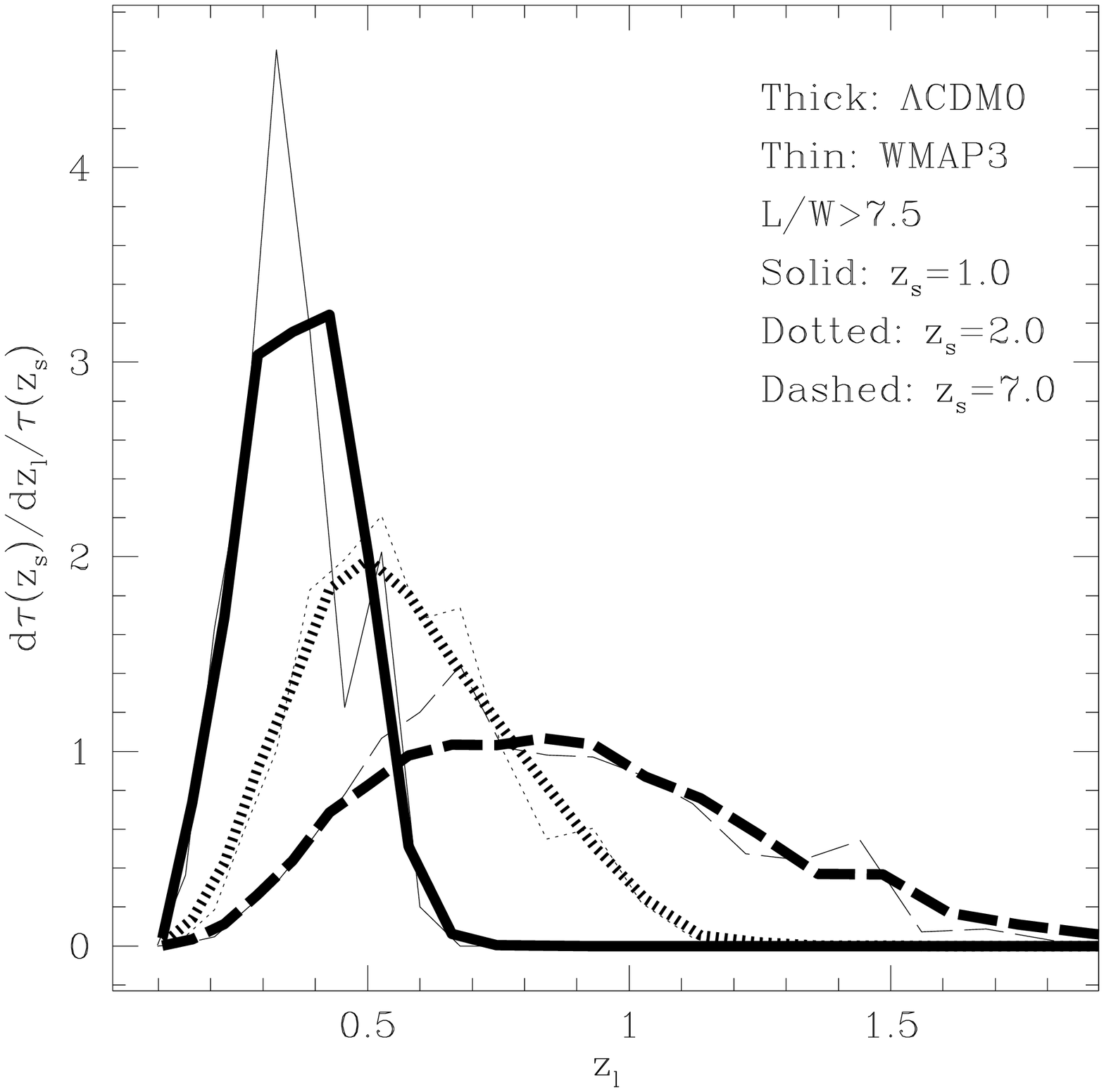}
}
\caption{The left panel shows the optical depth as a function
of source redshift in the $\LCDM$0 and WMAP3 models for giant
arcs with $L/W>7.5$ and $L/W>10$. The right panel shows the
differential probability distribution for the optical depth as a function of lens redshift
for $\zs=1, 2$ and 7 respectively. In both panels, the results for
the $\LCDM$0 and WMAP3 models are shown as thick and thin curves
respectively. 
} 
\label{fig:tau}
\end{figure}

The left panel of Fig. \ref{fig:massfunc} shows the mass functions in
the two cosmological models at redshift 0.3, the optimal lensing redshift
for a source at redshift 1 (see the right panel of Fig. \ref{fig:tau}). 
The mass functions are calculated following Sheth \& Tormen
(2002), which is based on the Press-Schechter (1974) formalism and the
ellipsoidal collapse model (Sheth, Mo \& Tormen 2001). The mass functions (shown as histograms) in
simulations clearly match the predictions well.
We can see that the number density of haloes in the WMAP3 model is always 
smaller than that in the $\LCDM$0 model on cluster mass scale.
For $M \sim 10^{14} h^{-1}M_\odot$, the abundance of haloes
is lower by a factor of 2.6 in the WMAP3 model compared with that in the
$\LCDM$0 model; for $M \sim 10^{15}h^{-1}M_\odot$, the reduction factor is $\sim 8.3$.
The lower abundance of clusters in the WMAP3 model implies that
ideally a larger simulation box is required to sample a similar number of
massive clusters as in the $\LCDM$0 model, a point we return to briefly in the discussion.

The right panel in Fig. \ref{fig:massfunc} shows how the predicted number density
of haloes above $10^{14}h^{-1} M_\odot$ evolves as a function of redshift in these two cosmologies.
The ratio of the number density of massive haloes in the WMAP3 model and
that in the $\LCDM$0 model as a function of redshift is well approximated (within 10\%) by
$0.4-0.2\,z$ for $z<1.5$. At redshift 0.3, 0.5 and 1, the number density of haloes
above $10^{14}h^{-1} M_\odot$ is about 1/3, 1/5 and 1/10 of that in the
$\LCDM$0 respectively. Notice that the cosmological distances and volumes also differ in the
three models, but the differences (which are accounted for in our calculations)
are small, ranging from 11\%  for the angular diameter distances and
6\% for the volume out to a source redshift of 2. These differences
are much smaller than the change in the cluster abundance.

As a consequence of the lower cluster abundance
in the WMAP3 model, one expects the lensing optical depth will also be much lower
compared with that in the $\LCDM$0 model. This is illustrated in
the left panel of Fig. \ref{fig:tau} which shows the optical depths as a function of the source
redshift in the $\LCDM$0 and WMAP3 models. First, we
notice that the optical depth is a strong function of the source
redshift, $\zs$, particularly when $\zs<3$, a trend first emphasized by \citet{Wambsganss04}.
Second, it is clear that the optical depth in the WMAP3 model is much
lower: compared with that in the $\LCDM$0 model, the optical depth
is reduced by more than a factor of $\sim 6$ for $L/W>7.5$ and
10 at all source redshifts. The right panel of Fig. \ref{fig:tau} shows the differential
probability distribution of the optical depth for giant arcs with $L/W>7.5$
as a function of the lens redshift for $\zs=1, 2$, and 7; the trends (not shown)
are similar for $L/W>10$. For a source at redshift 1, 2
and 7, the optimal lensing redshift is around 0.3, 0.5 and 0.75,
respectively, in the two cosmologies considered here. Note that the
normalised shapes are quite similar in the two cosmological models 
but the overall optical depth is much lower in the WMAP3 model.

\section{Discussions}

We have compared, using  high-resolution $N$-body simulations,  
the lensing probabilities of giant arcs in two
cosmological models: the old `concordance' model with $\sigma_8=0.9$,
and the model with parameters given by the three-year WMAP data. 
We find that the lensing probability decreases by a factor of $\sim 6$ 
in the 3-year WMAP model compared to that in the $\LCDM$0 concordance model. 
This decrease is largely a result of the much lower number density of 
massive haloes in the WMAP3 model (see Fig. 1).

There has been a long debate whether the number of observed giant arcs
is consistent with theoretical predictions. Early comparisons use
simple analytical models (\citealt{Kormann94, WH93, Wu96};
see also \citealt{Oguri03}), which were later shown to severely
under-estimate the lensing optical depth 
(Bartelmann \& Weiss 1994; Bartelmann, Steinmetz \& Weiss 1995; Meneghetti et al. 2003a; Torri et al. 2004).
Many recent studies use N-body simulations,
but with only dark matter (e.g. \citealt{Wambsganss04}; \citealt{Dalal04},
see \citealt{Puchwein05} for an exception). Below we discuss the
uncertainties in both observations and predictions, and examine
the consistency found in several previous works in the $\LCDM$0 cosmology.

\subsection{Consistency in the $\LCDM$0 cosmology}

Several previous studies concluded that the observational results 
of giant arcs are compatible with theoretical predictions (\citealt{Oguri03,
Wambsganss04, Dalal04}) in the $\LCDM$0 cosmology. 
However, there are uncertainties in such conclusion. 

The study by \citet{Oguri03} used
the axis ratio distributions in the tri-axial model of Jing \& Suto
(2002), and assumed a central profile, $\rho \propto r^{-1.5}$, 
that are steeper than the \citet{NFW97} form.  Gas cooling 
can steepen the halo density profile, but it also makes the central
mass distribution rounder. In this case, the shape distributions obtained
in dark matter-only simulations are not suitable for clusters 
where gas cooling  steepens the density profile dramatically. 
Since a shallower profile or a more spherical shape will reduce the 
giant arc cross-sections, it is unclear whether or not the conclusion 
of \citet{Oguri03} still holds if consistent models for density profile 
and halo shape are used. 
\citet{Wambsganss04} assumes that the $L/W$ ratio
can be approximated by the magnification. This assumption is valid 
for isothermal spheres, but not for real clusters which appear to be well
fit by the \citet{NFW97} profile (\citealt{vdM00, Com06, VF06}).
This assumption over-estimates the
optical depth by a factor of few (\citealt{Dalal04, Li05}). Furthermore,
they also assumed a slightly higher normalization, $\sigma_8 (0.95)$, 
which also increases the optical depth. The study by \citet{Dalal04}
assumes that arcs can be approximated as rectangles, 
which may overestimate the total cross-section by a factor of two 
compared to the perhaps more realistic assumption of elliptical arcs.
In summary, even in the $\LCDM$0 cosmology, the consistency between 
observations and predictions requires somewhat optimistic assumptions.
The WMAP3 model, which predicts an optical depth by a factor of 
$\sim 6$ smaller, makes it even harder to explain the observed giant 
arcs.

\subsection{Effects of baryons}

 In the real universe,
baryons account for roughly 20\% of the total mass, which
can cool (form stars) and sink toward the centres of clusters.
The radiative cooling likely has two effects: it will
increase the concentration of baryons at the center of clusters,
and at the same time, make the clusters more spherical (e.g.
\citealt{Dubinski94, Kaz04}). The former increases while the latter 
decreases the lensing cross-sections, and so the overall influence
depends on which effect dominates. 

Puchwein et al. (2005) studied the lensing cross-sections of clusters with $M \ga 10^{15}h^{-1}
M_\odot$ with different prescriptions of numerical viscousities. They concluded
that the higher concentrations due to baryons dominate and the baryons increase the
lensing cross-sections by a factor of $\sim 2$. However, they cautioned that
their clusters may suffer from overcooling, as the stellar density in
the core is larger than observed, and so the effect of baryons may be
over-estimated. Notice also that their simulations did not
include feedbacks from active galactic nuclei, which may 
further decrease the cooling of baryons at cluster centres 
and the corresponding lensing cross-sections. As the star formation
treatment in hydrodynamical simulations is uncertain,
\citet{Mene03b} adopted an alternate empirical approach. They studied the effect of
baryons by including a central cD galaxy
in clusters. They found that the increase in the lensing cross-section
due to cD galaxies is again quite moderate, by a factor of $ \la 2$. 
Such an  enhancement cannot compensate the reduction in the optical depth
due to the lower abundance of clusters in the WMAP3 model compared with the
$\LCDM$0 model.

\subsection{Uncertainties in the cluster and background source populations}

The cluster abundance (but not internal structures, particularly when star formation is
included) can now be reliably predicted from both numerical simulations and the 
extended Press-Schechter analytic formalism. In the $\LCDM$0 cosmology, 
various investigations, using similar assumptions of the background 
source population and definition of the length-to-width ratio,  
predict lensing optical depths that agree with one another within a 
factor of 1.5 (e.g. \citealt{Dalal04, Li05}; see \citealt{Li05} for 
a detailed discussion). We caution, however, that
even our $300 h^{-1}\mpc$ box size still appears to be somewhat too small
to sample the tail of the cluster mass function well, as is shown by
the large noise in the right panel of Fig. 2. We have
158 clusters with $M>10^{14} h^{-1}M_\odot$ (7 clusters with $M>3.8\times10^{14}h^{-1} M_\odot$).
At redshift $z=0.3$ for $\zs=3$, 50\% of the giant arc cross-section is
contributed by clusters with $M>3.8\times10^{14}h^{-1} M_\odot$. The
larger noise may also be partly due to our inadequate sampling of
merger events as we only dump the simulation data with a redshift
interval of $d z \approx 0.1$. As shown by a number of previous works
(Torri et al 2004; Hennawi et al. 2005; Fedeli et al. 2006), merger
events can boost the lensing cross-sections significantly. To examine
the importance of this effect, we ran a lower resolution simulation in
the WMAP3 model which evolves $N_{\rm DM}=512^3$ dark matter particles
in a box with sidelength of $600 h^{-1} \mpc$. Due to the larger volume,
 the new simulation
better samples the high mass tail of the cluster mass function and merger events. The
optical depth decreases by a factor of $\sim 2$.
The larger optical depth in our first WMAP3 simulation is partly
because it has relatively more massive
clusters than the larger simulation due to cosmic variance. Furthermore, the poorer resolution of
the larger simulation may have also reduced the optical depth, and
so the real change should be $\la 2$.

Substructures along the line of sight may be important for
the anomalous flux ratio problem (\citealt{Metcalf05}), but are unlikely to
be important for the lensing cross-sections as they are expected
to  contribute only a few per cent of the surface density.
Notice that we use a cube of $4h^{-1}\mpc$ to evaluate the lensing
cross-sections.  In \citet{Li05} the sidelength is chosen to be $2r_{\rm vir}$. These two choices give
almost identical optical depths (compare Fig. 2 with Fig.7 in
\citealt{Li05}). This demonstrates that the influence of matter
(including substructures) in the outer skirts of
clusters is not important.

A much larger uncertainty concerns the background source population,
including their ellipticity, size and redshift distributions.
While the former two appear to have modest
effects on the lensing cross-sections (e.g. \citealt{Oguri02, Li05}), 
the source redshift distribution
has quite dramatic effects on the lensing cross-section (Wambsganss et al. 2004), as can be
seen in Fig. \ref{fig:tau}. This is 
the biggest uncertainty in the predictions of giant arcs.
More redshift measurements of giant arcs will be particularly useful
to clarify the situation (see \citealt{Cov06} for a recent effort).

Currently the giant arc samples are still small. 
The largest dedicated search for giant arcs in X-ray clusters was performed by
Luppino et al. (1999) who found strong lensing in 8 out of 38 clusters. 
In the optical, Zaritsky \& Gonzalez (2003) found two giant
arcs using the Las Campanas Distant Cluster Survey between redshift 0.5
and 0.7 in an effective search area of 69 square degrees while
Gladders et al. (2003) found eight clusters with giant arcs
using the Red-Sequence Cluster Survey over $\sim 90$ square degrees.
The largest arc sample found from HST archives was presented by Sand et
al. (2005) with 116 arcs from 128 clusters, although its selection
function is likely heterogeneous. Clearly the current giant arc samples
are somewhat limited, although future weak lensing surveys (e.g. \citealt{Wittman06}) may  
yield a large number of giant arcs as a by-product. A combined analysis of
strong and weak lensing will be particularly useful for constraining
$\sigma_8$ and the cluster inner mass profiles.

Until larger arc samples become available and our understanding
of the background source population is improved, it is difficult to 
reach firm conclusions concerning the consistency of the WMAP three-year model with
observations. Nevertheless, it appears difficult to reconcile
the giant arc statistics with the low central $\sigma_8$ value (0.74) preferred
by the WMAP three-year data.

\section*{Acknowledgment}
We thank Volker Springel for providing the code GADGET2 and the referee
for insightful comments. This work is supported by grants from 
NSFC (No. 10373012, 10533030) and Shanghai Key Projects in Basic research
(No. 04JC14079 and 05XD14019). HJM, LG and SM acknowledge travel support from
the Chinese Academy of Sciences. 
The simulations were performed at the Shanghai Supercomputer Center.

\end{document}